\pdfoutput=1
\documentclass[11pt,twoside,a4paper,cmspaper,final,collab]{cms-tdr}
\def\svnVersion{e6a1bdd}\def\svnDate{2025/09/22}\def\cmsCernNoTag{CERN-EP-2025-108}\def\cmsCernDate{\today}\def\cmsMessage{Published in Physical Review Letters as \href{http://dx.doi.org/10.1103/zc76-rgcp}{\doi{10.1103/zc76-rgcp}.}}
\begin{document}\cmsNoteHeader{BPH-23-008}

\newcommand{\dzmm}      {\ensuremath{\PDz\to\PGmp\PGmm}\xspace}
\newcommand{\dzpipi}    {\ensuremath{\PDz\to\PGpp\PGpm}\xspace}
\newcommand{\kspipi}    {\ensuremath{\PKzS\to\PGpp\PGpm}\xspace}
\newcommand{\dzkpi}     {\ensuremath{\PDz\to \PKm\PGpp}\xspace}
\newcommand{\dzpipimm}  {\ensuremath{\PDz\to\PGpp\PGpm\to\PGmp\PGn\PGmm\PAGn}\xspace}
\newcommand{\dstartodz} {\ensuremath{\PDstp\to\PDz\PGpp}\xspace}
\newcommand{\dstarpipi} {\ensuremath{\PDz\to\PGp\PGp}\xspace}
\newcommand{\dstarpipimm} {\ensuremath{\PDz\to\PGp\PGp\to\PGm\PGn\PGm\PGn}\xspace}
\newcommand{\dstarmm}   {\ensuremath{\PDz\to\PGm\PGm}\xspace}
\newcommand{\dstarpimunu} {\ensuremath{\PDz\to\PGp\PGm\PGn}\xspace}
\newcommand{\dzpimunu}  {\ensuremath{\PDz\to \PGpm\PGmp\PGn}\xspace}
\newcommand{\m}        {\ensuremath{m(\PDz)}\xspace}
\newcommand{\dm}        {\ensuremath{\Delta m}\xspace}
\newcommand{\mvad}      {\ensuremath{d_{\text{MVA}}}\xspace}
\newcommand{\mvadcorrection}      {\ensuremath{d_{\text{MVA, cor}}}\xspace}
\newcommand{\mpipi}     {\ensuremath{m_{\PGp\PGp}}\xspace}
\newcommand{\mmm}       {\ensuremath{m_{\PGm\PGm}}\xspace}
\newcommand{\cbf}       {\ensuremath{\mathcal{B}}}
\newcommand{\eps}       {\varepsilon\xspace}
\newcommand{\phikk}     {\ensuremath{\PGfP{1020} \to \PKp\PKm}\xspace}
\newcommand{\bdmm}      {\ensuremath{\Bd \to \MM}\xspace}
\newcommand{\bsmm}      {\ensuremath{\Bs \to\MM}\xspace}
\newcommand{\Bd}        {\PBz}
\newcommand{\Bs}        {\PBzs}
\newlength\cmsTabSkip\setlength{\cmsTabSkip}{1ex}
\ifthenelse{\boolean{cms@external}}{\providecommand{\cmsLeft}{upper\xspace}}{\providecommand{\cmsLeft}{left\xspace}}
\ifthenelse{\boolean{cms@external}}{\providecommand{\cmsRight}{lower\xspace}}{\providecommand{\cmsRight}{right\xspace}}
\ifthenelse{\boolean{cms@external}}{\providecommand{\cmsTopRight}{middle\xspace}}{\providecommand{\cmsTopRight}{right\xspace}}
\ifthenelse{\boolean{cms@external}}{\providecommand{\cmsTwoUpper}{top and middle plots\xspace}}{\providecommand{\cmsTwoUpper}{upper plots\xspace}}
\ifthenelse{\boolean{cms@external}}{\providecommand{\largePlotWidth}{0.48\textwidth}}{\providecommand{\largePlotWidth}{0.65\textwidth}}
\providecommand{\cmsTable}[1]{\resizebox{\textwidth}{!}{#1}}
\ifthenelse{\boolean{cms@external}}{\providecommand{\tableCMS}[1]{\resizebox{\columnwidth}{!}{#1}}}{\providecommand{\tableCMS}[1]{#1}}
\ifthenelse{\boolean{cms@external}}
{
\newcommand{\appname}{End Matter} 
\newcommand{\appref}{the \appname} 
\def\appendixname{\relax}}
{
\newcommand{\appname}{\appendixname}
\newcommand{\appref}{\appname~\ref{end_matter_a}}
}

\cmsNoteHeader{BPH-23-008}
\title{Search for the rare decay \texorpdfstring{$\PDz\to\PGmp\PGmm$}{D0 to mu+ mu-} in proton-proton collisions at \texorpdfstring{$\sqrt{s} = 13.6\TeV$}{sqrt(s) = 13.6 TeV}}

\author*[cern]{The CMS Collaboration}

\date{\today}

\abstract{A search for the rare decay $\PDz\to\PGmp\PGmm$ is  reported using proton-proton collision events at $\sqrt{s} = 13.6\TeV$  collected by the CMS detector in 2022--2023, corresponding to an  integrated luminosity of 64.5\fbinv. This is the first analysis  to use a newly developed inclusive dimuon trigger, expanding the scope of the CMS flavor physics program. The search uses \PDz mesons obtained from $\PDstp\to \PDz \PGpp$ decays. No significant excess is observed. A limit on the branching fraction of $\mathcal{B}(\PDz\to\PGmp\PGmm) < 2.4 \times 10^{-9}$ at 95\%  confidence level is set. This is the most stringent upper limit set on any flavor changing neutral current decay in the charm sector. }

\hypersetup{%
pdfauthor={CMS Collaboration},%
pdftitle={Search for the rare decay D0 to mu+ mu- in proton-proton collisions at sqrt(s) = 13.6 TeV},%
pdfsubject={CMS},%
pdfkeywords={CMS, charm physics, flavor physics, rare decays}}

\maketitle

Studies of rare decays of hadrons have long been considered one of the most promising avenues for discovering physics beyond the standard model (SM).
Contributions from new physics effects are more easily detected in rare decays, 
as the smaller SM contributions make deviations more apparent. 
Decays mediated by flavor changing neutral currents, which are forbidden at 
tree level in the SM, have been studied in various experiments.
These studies have
primarily focused on rare decays of bottom and strange hadrons involving $\PQb\to
\PQs$ and $\PQs\to \PQd$ transitions~\cite{PDG2024,HFLAV2024}.
Rare decays of charmed hadrons via the $\PQc\to \PQu$ process have received less attention.

An important distinction between rare decays of charm hadrons and those of bottom or 
strange hadrons is that the loop contributions in charm decays are mediated by
lighter quarks. This leads to substantial long-distance SM contributions (arising from nonperturbative hadronic effects), which are
challenging to calculate, motivating searches where the beyond SM predictions substantially exceed those of the SM.
The rare decay of the \PDz meson into two muons is one such case. The SM prediction for the branching fraction $\cbf(\dzmm)$ is around
$3\times10^{-13}$~\cite{Burdman:2001tf}. New physics models~\cite{PhysRevD.79.114030,LQ3,LQ4,Burdman:2001tf}, including those involving leptoquarks, R-parity violating supersymmetry, and extra fermions or gauge bosons, predict enhancements to this decay rate, which could be accessible by experiments at the LHC. 
This decay also provides stringent constraints on the Wilson coefficients $C_{S,P,10}$ in the effective field theory framework~\cite{eft1,eft2,eft3}.
The most sensitive experimental search
to date, conducted by the LHCb Collaboration, set an upper limit on $\cbf(\dzmm)$ at
$3.1~(3.5)\times10^{-9}$ at 90~(95)\% confidence level (\CL)~\cite{LHCb:2022jaa}, which still allows for substantial new physics contributions above the SM.

This Letter reports a search for \dzmm decays by the CMS experiment based on proton-proton (pp) collision data collected in 2022--2023 at a center-of-mass energy of 13.6 TeV, corresponding to an integrated luminosity of 64.5\fbinv. This search is enabled by a new inclusive dimuon trigger~\cite{parkingtrigger}, which significantly broadens the scope of the CMS flavor physics program and is used here for the first time. Tabulated results are provided in the HEPData record for this analysis~\cite{hepdata}.

One of the main challenges in the search for \dzmm decays is suppressing the combinatorial background, which dominates the overall background.
To do so, we look for the signal in the decays of the $\PDst(2010)^{+}$ meson, i.e., \dstartodz.  In this Letter, charge-conjugated decay modes are implicitly included, and \PDz refers to \PDz mesons from a \PDstp decay unless otherwise indicated. Although this approach reduces the event yield by about a factor of four relative to inclusive $\PDz$ meson production,  the requirement of a pion from the \PDstp decay significantly improves vertex constraints and reduces the dominant background by two orders of magnitude. To extract the signal, a two-dimensional (2D) unbinned maximum likelihood (UML) fit is performed on the \PDz candidate mass \m and the mass difference between the $\PDstp$ and the daughter $\PDz$ mesons $\dm=m(\PDstp)-\m$.  
The small energy release in the \PDstp meson decay results in a narrow peak in the \dm spectrum, which can be used to greatly suppress the background. The decay kinematics result in the pion from the $\PDstp$ meson decay tending to have low momentum, which is therefore referred to as a soft pion.

The branching fraction of interest is calculated with respect to a normalization decay channel, \dzpipi, using
\begin{equation}\label{eq:br}
\cbf(\dzmm) = \cbf(\dzpipi)\frac{N_{\dstarmm}}{N_{\dstarpipi}} \frac{\eps_{\dstarpipi}}{\eps_{\dstarmm}}, 
\end{equation}
where $N_{\mathrm{x}}$ is the fitted yield for the decay $\mathrm{x} \in (\PDz\to\PGm\PGm, \PDz\to\PGp\PGp)$, and $\eps_{\mathrm{x}}$ is the corresponding full selection efficiency. 
Another possible normalization channel is \dzkpi. While the \dzkpi branching fraction is larger than that for the \dzpipi decay, the kinematics are more similar for \dzpipi, and therefore \dzkpi decays are only used as a cross-check. 
Measuring the branching fraction relative to the normalization channel eliminates uncertainties related to the production of $\PDstp$ mesons, including the fraction of $\PDstp$ mesons produced directly or from $\PQb$ hadron decays. In addition, many systematic uncertainties related to the reconstruction and selection efficiency are reduced.

The CMS apparatus~\cite{CMS:2008xjf,CMS:2023gfb} is a multipurpose, nearly hermetic detector, designed to trigger on~\cite{CMS:2020cmk,CMS:2016ngn,CMS:2024aqx} and identify electrons, muons, photons, and hadrons~\cite{CMS:2020uim,CMS:2018rym,CMS:2014pgm}. A global algorithm~\cite{CMS:2017yfk} aims to reconstruct all individual particles in an event, combining information provided by the all-silicon inner tracker and by the crystal electromagnetic and brass-scintillator hadron calorimeters, operating inside a 3.8\unit{T} superconducting solenoid, with data from gas-ionization muon detectors embedded in the flux-return yoke outside the solenoid. 

Simulated events are used to evaluate the signal and normalization efficiencies, the detector response, and the background yields. The simulated
events are generated with the SoftQCD:nonDiffractive option in \PYTHIA\ 8.212~\cite{Sjostrand:2014zea} using
the CP5 underlying event tune~\cite{CMS:2019csb} and are propagated through the CMS
detector model using the \GEANTfour~\cite{Agostinelli:2002hh} package. The
decays of charm mesons are simulated using the \EVTGEN 1.3.0~\cite{Lange:2001uf}
program and final-state photon radiation is described using \PHOTOS
3.56~\cite{Davidson:2010ew}. Additional pp interactions per bunch crossing (pileup) are added to the simulated samples with a multiplicity distribution matching the data.

Events of interest are selected using a two-tiered trigger system. The first level (L1), composed of custom hardware processors, uses information from the calorimeters and muon detectors to select events at a rate of around 100\unit{kHz} within a fixed latency of 4\mus~\cite{CMS:2020cmk}. The high-level trigger (HLT) consists of a farm of processors running a version of the full event reconstruction software optimized for fast processing, and reduces the event rate to about 5\unit{kHz} before data storage~\cite{CMS:2016ngn,CMS:2024aqx}.
For LHC Run 3, which began in 2022, a new trigger strategy was designed and deployed using inclusive dilepton trigger algorithms with low transverse momentum (\pt) thresholds and a data parking technique~\cite{parkingtrigger}, in which the data stream is saved for delayed reconstruction. This provides a high efficiency for hadron decays with two leptons in the final state, which are central to the CMS flavor physics program. The trigger used to collect the signal events required one muon with $\pt>3\GeV$ and another with $\pt>4\GeV$. The normalization channel events are collected with a zero bias trigger that selects one in every $\approx$1.4 million colliding bunches.  

The offline selection begins with the reconstruction of \dzmm and \dzpipi
candidates, with more stringent criteria than the trigger requirements. 
The selection criteria for the two channels are kept as consistent as possible to minimize the systematic uncertainty in the efficiency ratio. 
The charged particle tracks used to reconstruct the $\PDz$ candidates must pass high purity criteria~\cite{CMS:2014pgm}. For the signal channel, each muon candidate must also have a reconstructed track in the tracker that matches segments in the barrel or endcap muon detectors, and a compatible track in the muon system, while passing a loose muon identification requirement~\cite{CMS:2018rym}, chosen through sensitivity optimization.
The tracks of the $\PDz$ candidates are refitted to a common vertex~\cite{Prokofiev:2005zz}. 
The absolute distance between the closest point on the $\PDz$ trajectory and a reconstructed pp collision vertex in 3D space is defined as the impact parameter.
The primary vertex (PV) is taken to be the pp collision vertex with the smallest impact parameter to the $\PDz$ candidate.
The angle between the dimuon momentum and the line connecting the PV to the \PDz vertex is defined as the pointing angle.
Soft pion candidates are selected from the charged tracks compatible with this PV. 
To improve the $\dm$ resolution, the PV is refitted using only the soft pion and the tracks consistent with emerging from the PV, and this vertex is used as a constraint on the soft pion trajectory to refine its momentum. This refitted PV serves as the \PDstp candidate vertex. 
Finally, the $\PDz$ candidate is required to have decay products, either muons or pions, with $\pt>4\GeV$ and pseudorapidity $\abs{\eta}<2.4$, a reconstructed invariant mass in the range 1.81--1.94\GeV, a $\PDz$ vertex fit $p$-value $>$0.01, a significance of the \PDz vertex distance from the PV in 3D greater than 3, and a pointing angle in 3D less than 0.1. The \PDstp candidate is required to have a vertex fit $p$-value $>$0.1 and \dm in the range 0.14--0.15\GeV.  

A multivariate analysis is employed using the XGBOOST
package~\cite{Chen:2016:XST:2939672.2939785} to construct a boosted decision tree discriminator (\mvad). 
The input features include the \pt of the muons and the soft pion, the \PDz candidate's pointing angle, vertex distance, and impact parameter relative to the PV, vertex fit $p$-values of the \PDz and \PDstp candidates, and the \PDz mass resolution divided by the reconstructed \PDz mass.
The signal events used in the training are obtained from simulated \dzmm events, where the \PDstp meson can be produced either directly or via \PQb-hadron decays, as both production modes contribute to the potential signal. The background events used in training are from a data sideband region defined by $0.150<\dm<0.155\GeV$ and $1.81<\mmm<2.45\GeV$. The $k$-folding technique with $k=5$ is applied for the MVA training~\cite{kfold}. 
Signal and normalization channel candidates are selected by applying the same \mvad requirement, which greatly reduces systematic uncertainties in the efficiency ratio. The \mvad threshold value is optimized to give the most stringent expected limit on $\cbf(\dzmm)$, improving it by 60\% relative to the selection without the \mvad requirement.  
A small difference in selection efficiency between the two channels, determined from simulated events, is incorporated into the efficiency correction. 
After selection, about 1.5\% of the events in each channel contain multiple candidates, with no requirement made to select a single candidate.

\begin{figure*}[htp]
  \centering
    \includegraphics[width=0.45\textwidth]{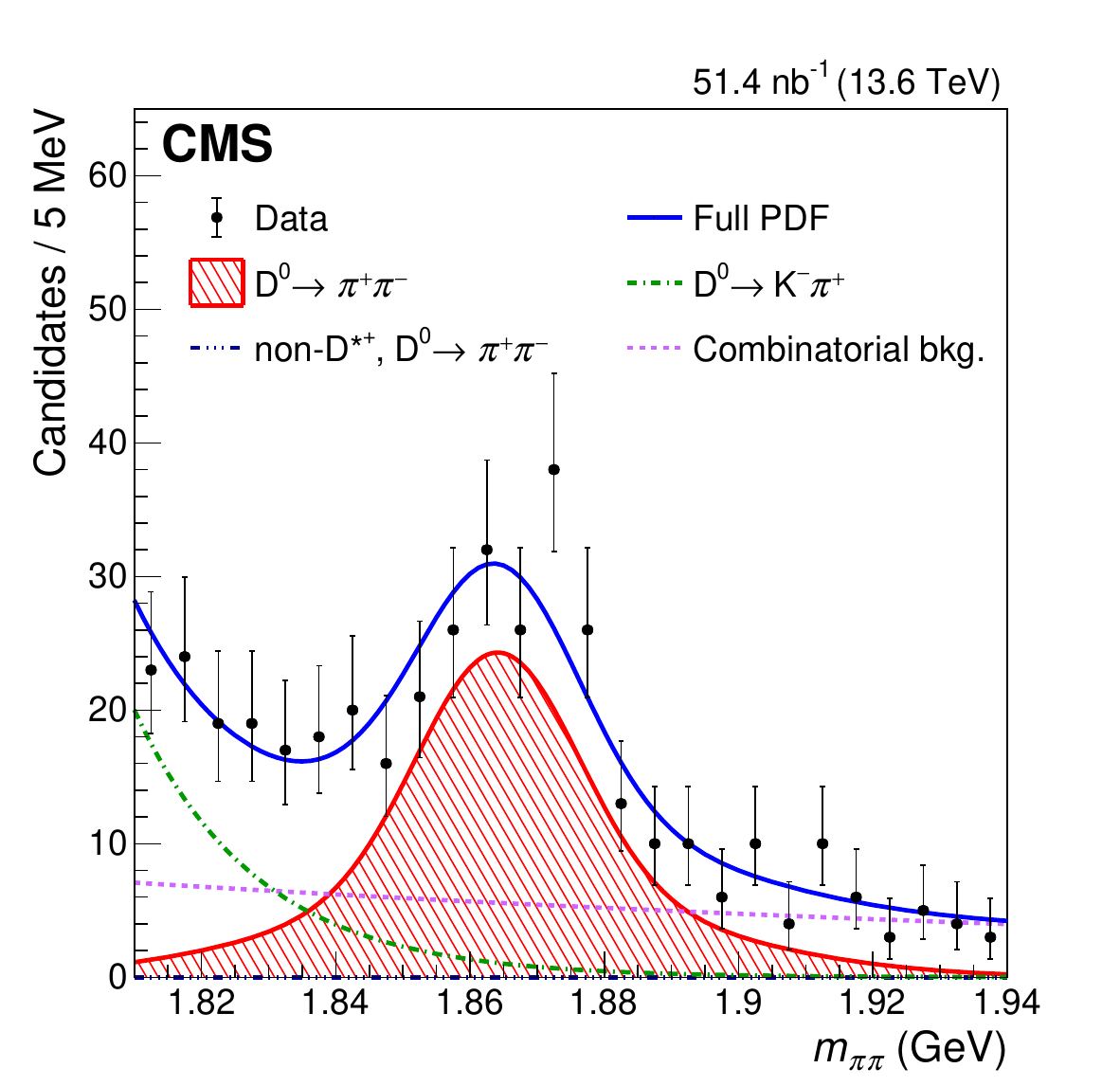}
    \includegraphics[width=0.45\textwidth]{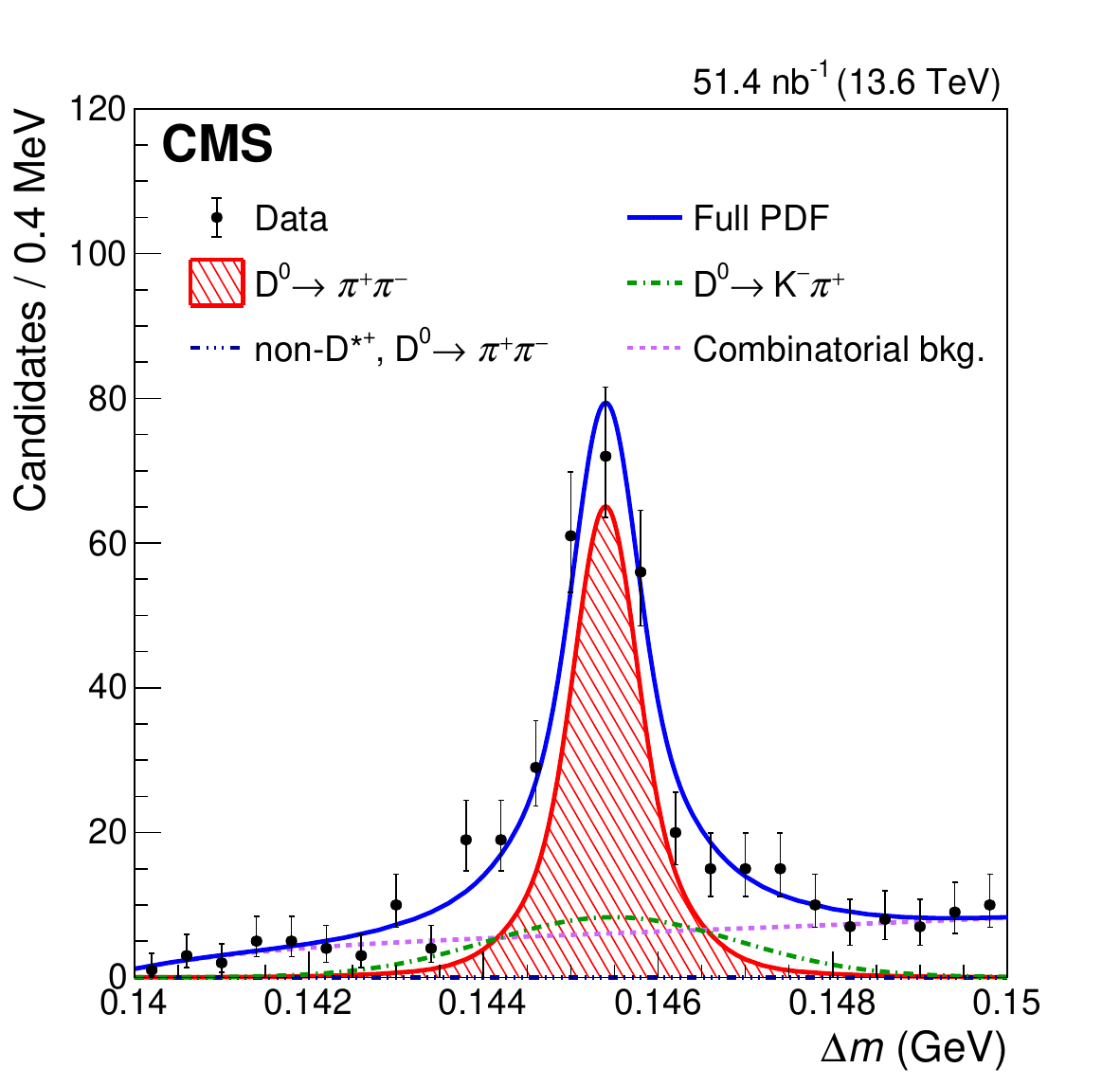}\\
\caption{The distributions of the dipion invariant mass \mpipi (left) and $\PDst$--$\PDz$ mass difference \dm (right) along with the associated projections of the full fit (solid curve), the signal contribution (hatched area), and the background contributions (other curves).}
  \label{fig:dpipi_fit_zero_bias_afterMva}
\end{figure*}

The normalization channel yield is extracted from a 2D UML fit to the dipion invariant mass (\mpipi) and \dm, as shown in Fig.~\ref{fig:dpipi_fit_zero_bias_afterMva}.
The 2D probability density functions (PDFs) for signal and background components are constructed as the product of two 1D PDFs, as simulation studies showed no correlation between the two variables. The signal PDFs are a sum of two (three) Gaussian functions for \mpipi (\dm), with a common mean. The background from the \dzkpi decays is modeled using an exponential function for \mpipi and a Gaussian function for \dm. 
The combinatorial background is modeled with an exponential function in \mpipi and the following function for \dm: 
  \begin{equation}
    \left(1 - \re^{-\frac{\dm - m_{\pi}}{C}} \right)
    \left(\frac{\dm}{m_{\pi}}\right)^A + B
    \left(\frac{\dm}{m_{\pi}} - 1 \right), \label{eq:pdf}
  \end{equation}
where $A$, $B$, and $C$ are fit parameters. The parameter $B$ is fixed to zero, whereas it is left unconstrained in the signal channel fit. For non-\PDstp backgrounds (where the \PDz meson decays to $\PKm\PGpp$ or $\PGpp\PGpm$ and does not originate from a \PDstp meson), the \mpipi distribution is modeled similarly to that of the corresponding \PDstp fit components, while \dm is described by Eq.~\eqref{eq:pdf}, again with $B=0$. In the fit, the yields, the parameters of the combinatorial background PDF, and the \mpipi exponential decay constant for the \dzkpi background are free parameters, with the remaining parameters fixed from prefits to simulated and dimuon-triggered events, which contain many more events than the zero bias sample. The signal shape parameters for both \mpipi and \dm are initially established from fits to simulated events, with the means and widths multiplied by correction factors. The correction factors are obtained from a fit to the dimuon-triggered events to correct the simulation, and the width correction factors are additionally corrected to account for differences between the zero bias and dimuon triggered event samples by comparing the widths of \dzkpi events in the two samples. The remaining parameters are obtained from fits to the dimuon-triggered events. 
 Fits to the dimuon-triggered sample found a small contribution from non-\PDstp, \dzpipi, while the contribution from non-\PDstp, \dzkpi was consistent with zero. Therefore, only the former contribution was included in the fit to the zero bias sample, with a resultant yield consistent with zero. 
 The fitted yield of \dzpipi decay is $195\pm17\stat$, from an effective integrated luminosity of 51.4~${\mathrm{nb}}^{-1}$.

\begin{figure*}[ht]
\centering
\includegraphics[width=0.45\textwidth]{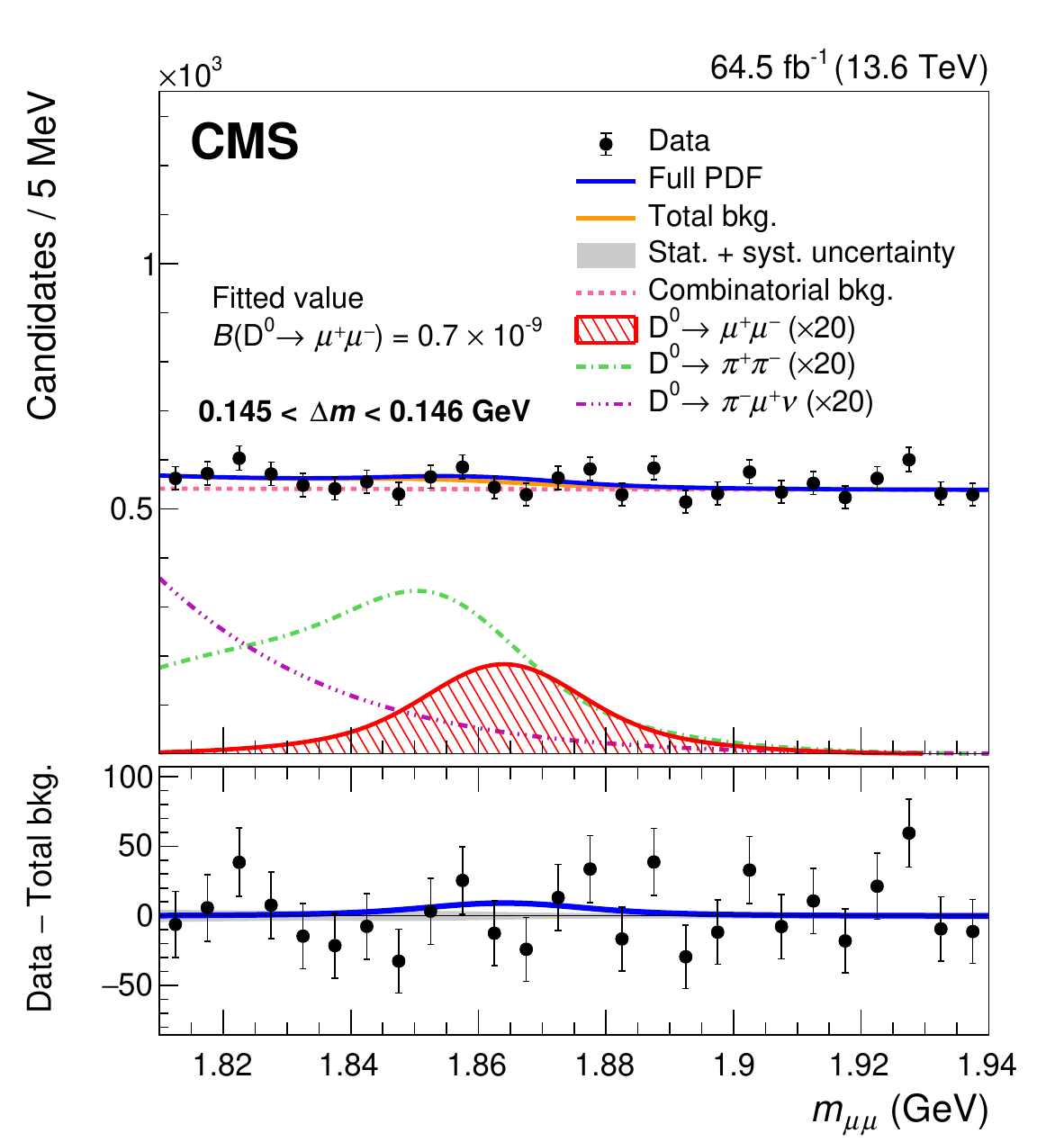}
\includegraphics[width=0.45\textwidth]{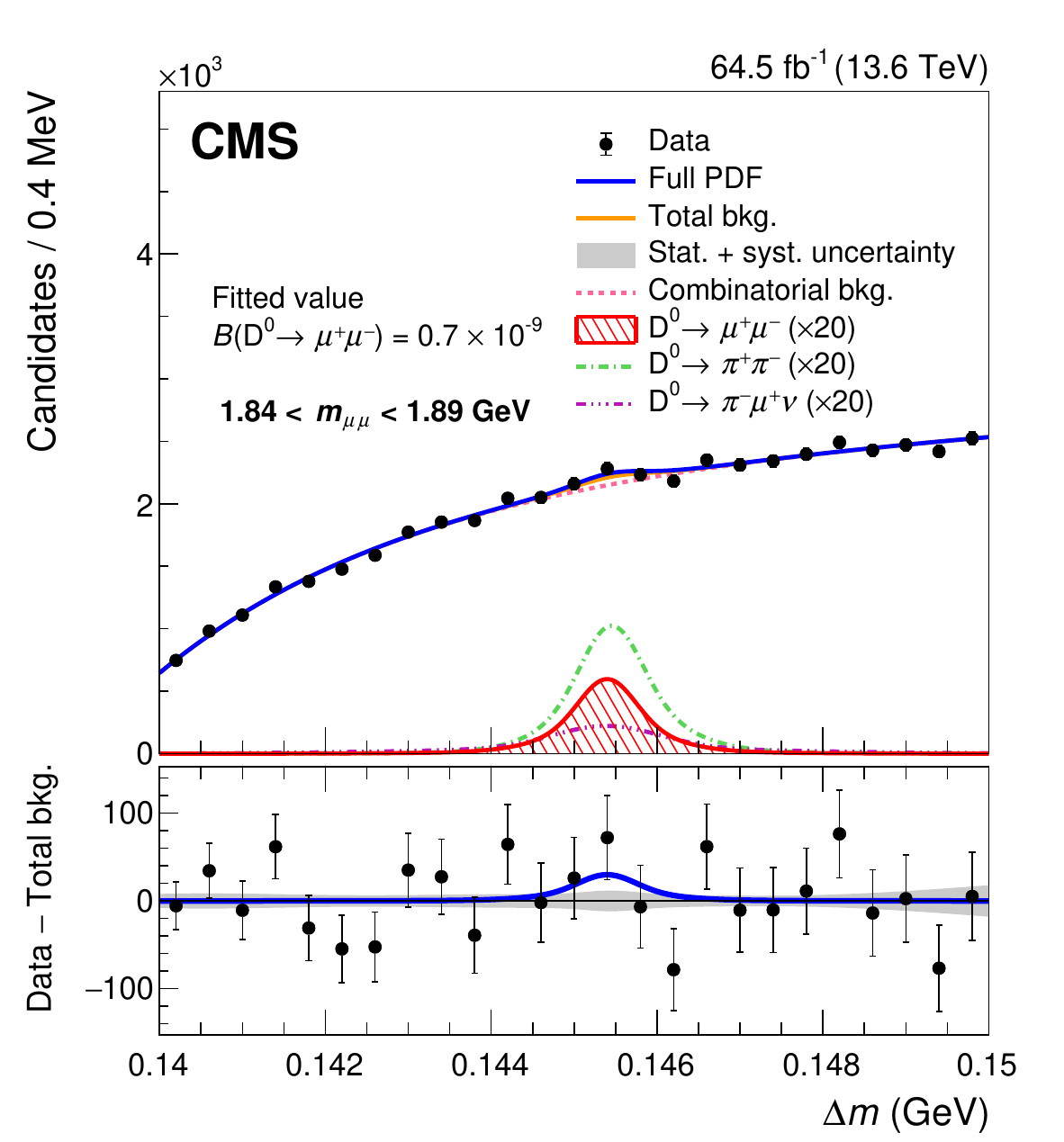}
    \caption{The distributions of \mmm (left) and \dm (right) of the fit with the requirements $0.145<\dm<0.146\GeV$ and $1.84<\mmm<1.89\GeV$, respectively, along with the associated projections of the full fit (solid curve), signal contribution (hatched area), and background contributions (other curves). The \PDz meson components are scaled up by 20 in the upper panel. The lower panel shows the data and the fit result after subtracting the total background component of the fit. The gray error band represents the statistical and systematic uncertainties in the total background component.}
    \label{fig:projection_cut_unblind2}
\end{figure*}

The signal channel yield is extracted in a similar way, with the dimuon invariant mass \mmm replacing \mpipi. The \mmm distribution for $\dm \in [0.145,0.146]\GeV$ and the \dm distribution for $\mmm \in [1.84, 1.89]\GeV$ are shown in Fig.~\ref{fig:projection_cut_unblind2}. Four components contribute to this channel: the signal \dzmm decay, backgrounds from the $\dzpipi$ and  $\dzpimunu$ processes in which charged pions are misidentified as muons, and the combinatorial background. The contribution of non-\PDstp backgrounds that could peak in the \mmm distribution are found to be negligible.

The signal model for \dzmm uses the same PDF form as in the normalization channel: a 2D PDF consisting of two (three) Gaussian functions for \mmm (\dm), with a common mean. The two backgrounds from \PDz decays are similarly modeled, each using two Gaussians for \mmm with different means and three Gaussians for \dm with a common mean. The model parameters for these three components are determined from simulated events. The means and widths of the corresponding signal model parameters are multiplied by the correction factors derived for the normalization channel. The \mmm combinatorial background shape is modeled with either a first-order Bernstein polynomial~\cite{ref:bernstein}, a power law, or an exponential function, using the discrete profiling method~\cite{Dauncey:2014xga}, which allows the choice of function to be determined during the fit minimization process with a discrete nuisance parameter. The \dm combinatorial background shape is modeled using Eq.~\eqref{eq:pdf}, with no constraints on the fit parameters, including $B$. While the background is dominated by combinatorial processes, the two \PDz background sources are important to estimate, as both exhibit peaks in the \dm distribution at the signal location and the \dzpipi also peaks in \mmm, albeit slightly shifted from the signal. Both of these sources are normalized using the measured \dzpipi yield in the normalization channel and by applying the appropriate scaling from luminosity, branching fractions, efficiencies, and muon misidentification to determine the background yield. The details of this normalization are found in \appref.

Most of the systematic effects encountered in the data analysis cancel in the ratio of the signal and normalization channel yield measurements. Systematic uncertainties that do not cancel are incorporated into the UML fit as nuisance parameters.
The dimuon trigger efficiencies are measured with respect to high-efficiency reference triggers using control samples with muon kinematics similar to the signal samples.
The resulting data-to-simulation ratio is 0.943 with an uncertainty of 0.7\%, accounting for statistical uncertainties and variations across data-taking periods and reference triggers.
The muon reconstruction and identification efficiency correction is measured using the ``tag-and-probe'' method~\cite{ref:TnP} with  $\PJGy\to\MM$ decays and found to be consistent with unity in the kinematic region relevant to the analysis. As such, no correction is applied but the corresponding 1\% uncertainty is included for each muon.
The track reconstruction efficiency has a 2.3\% uncertainty per track~\cite{DP-2022-012}. The difference in pileup between the zero bias and dimuon triggers introduces a 1\% uncertainty in the selection efficiency ratio. The statistical uncertainty in the yield of \dzpipi events contributes as a systematic uncertainty for the normalization of the signal (Eq.~\eqref{eq:br}) and the two \PDz backgrounds. The limited number of simulated events results in an efficiency uncertainty. 
The \mvad selection efficiencies for the signal and normalization channels differ by 1\%. This correction is made and uncertainties of 1.2\% and 2\% are assigned for the \dzmm signal and \dzpipi background, respectively, accounting for the deviation from unity and the statistical uncertainty from the limited number of simulated events. A disagreement of similar size is found for the \dzpimunu background but no correction or uncertainty is applied as it is much smaller than the efficiency uncertainty of 12\%.
The muon misidentification rate systematic uncertainty of 14\% includes statistical and systematic uncertainties related to the measurement in $\kspipi$ decays as well as the treatment of decay-in-flight and non-decay-in-flight sources in simulation.
The fit bias uncertainty is 1\%, determined by the difference between the input and output of UML fits using pseudo-experiments with an injected \dzmm signal. The variation from using alternative modeling for signal and background components is examined and found to be negligible. Table~\ref{tab:systematics_summary} presents a summary of all systematic uncertainties, including the \dzpipi and \dzpimunu branching fraction uncertainties~\cite{PDG2024}, with additional information provided in \appref.

\begin{table}[!htb]
  \centering
  \caption{Summary of systematic uncertainties for the \dzmm branching fraction measurement with their corresponding contributions in the signal channel.}
    \label{tab:systematics_summary}
    \begin{scotch}{lccc}
      Source                          & \dstarmm  & \dstarpipi & \dstarpimunu \\
      \hline
      Trigger efficiency   &  0.7\% & 0.7\% & 0.7\%  \\
      Muon efficiency       &  2\% &  \NA & 1\% \\
      Tracking efficiency  & 4.6\% & 4.6\% & 4.6\%  \\
      Pileup                           &  1\% &  1\% & 1\% \\
      \dzpipi yield            & 8.7\% & 8.7\% & 8.7\% \\
      Efficiency  &  0.2\%  & 0.6\%  & 12\% \\
      \mvad correction        &  1.2\% & 2.0\% & \NA \\
      $\cbf(\dzpipi)$                    & 1.7\%  & \NA & 1.7\% \\
      $\cbf(\dzpimunu)$                  &  \NA  &  \NA &  4.5\% \\
      Fit bias                      & 1\%  &  \NA & \NA \\
      Misidentification rate  & \NA & 28\% & 14\% \\
    \end{scotch}
\end{table}

The results are obtained using the \textsc{Combine} tool~\cite{combinetool}.  The nuisance parameters associated with the tracking efficiency, muon efficiency, pileup, fit bias, and \mvad correction are constrained with log-normal PDFs, while the ones associated with the efficiency, trigger efficiency, muon misidentification rate, input branching fractions, and the \dzpipi yield are constrained with Gaussian PDFs. 

No obvious excess of events is observed in the data. 
The fit returns a signal yield of $100\pm 120$, resulting in a branching fraction of $\cbf(\dzmm)= (0.7\pm 0.9) \times 10^{-9}$. Additional information on signal and background yields can be found in \appref.
The observed significance, evaluated from the log-likelihood difference, is 0.8 standard deviations. An upper limit is set using the asymptotic \CLs method~\cite{Junk,Read}:
\begin{equation*}
\cbf(\dzmm) < 2.1\,(2.4) \times 10^{-9}\ \text{at 90\,(95)\% \CL}.
\end{equation*} 
The observed limit is higher than the expected limit from background-only pseudo-experiments, which is $\cbf(\dzmm) < 1.5\,(1.9)\times 10^{-9}$ at 90\,(95)\% \CL.

The fit and likelihood scan are repeated with each of the three functions describing the combinatorial background in \mmm, without using the discrete profiling method, and return consistent results. The result using the alternative normalization channel $\dzkpi$ is consistent with the nominal result.

In summary, a search for \dzmm decays by the CMS experiment, using proton-proton collision data at $\sqrt{s} = 13.6\TeV$ corresponding to an integrated luminosity of 64.5\fbinv, is presented. 
No significant excess above the fitted background is observed. An upper limit of $\cbf(\dzmm) < 2.4 \times 10^{-9}$ is set at 95\% confidence level. This search is the most sensitive to date and provides a significant improvement over the previous best result~\cite{LHCb:2022jaa}, setting the most stringent limit on flavor changing neutral currents in the charm sector. It can be used to set constraints on scenarios that modify $\cbf(\dzmm)$.  The search is made possible by a newly developed inclusive dimuon trigger and represents its first application. The result demonstrates the benefits of this trigger for flavor physics measurements and its potential to enable opportunities for a wide range of studies involving low-mass muon pairs.

\begin{acknowledgments}
We congratulate our colleagues in the CERN accelerator departments for the excellent performance of the LHC and thank the technical and administrative staffs at CERN and at other CMS institutes for their contributions to the success of the CMS effort. In addition, we gratefully acknowledge the computing centres and personnel of the Worldwide LHC Computing Grid and other centres for delivering so effectively the computing infrastructure essential to our analyses. Finally, we acknowledge the enduring support for the construction and operation of the LHC, the CMS detector, and the supporting computing infrastructure provided by the following funding agencies: SC (Armenia), BMBWF and FWF (Austria); FNRS and FWO (Belgium); CNPq, CAPES, FAPERJ, FAPERGS, and FAPESP (Brazil); MES and BNSF (Bulgaria); CERN; CAS, MoST, and NSFC (China); MINCIENCIAS (Colombia); MSES and CSF (Croatia); RIF (Cyprus); SENESCYT (Ecuador); ERC PRG, RVTT3 and MoER TK202 (Estonia); Academy of Finland, MEC, and HIP (Finland); CEA and CNRS/IN2P3 (France); SRNSF (Georgia); BMBF, DFG, and HGF (Germany); GSRI (Greece); NKFIH (Hungary); DAE and DST (India); IPM (Iran); SFI (Ireland); INFN (Italy); MSIP and NRF (Republic of Korea); MES (Latvia); LMTLT (Lithuania); MOE and UM (Malaysia); BUAP, CINVESTAV, CONACYT, LNS, SEP, and UASLP-FAI (Mexico); MOS (Montenegro); MBIE (New Zealand); PAEC (Pakistan); MES and NSC (Poland); FCT (Portugal); JINR (Dubna); MON, RosAtom, RAS, RFBR, and NRC KI (Russia); MESTD (Serbia); MICIU/AEI and PCTI (Spain); MOSTR (Sri Lanka); Swiss Funding Agencies (Switzerland); MST (Taipei); MHESI and NSTDA (Thailand); TUBITAK and TENMAK (Turkey); NASU (Ukraine); STFC (United Kingdom); DOE and NSF (USA).
\end{acknowledgments}

\bibliography{auto_generated}

\providecommand{\href}[2]{#2}\begingroup\raggedright\begin{thebibliography}{10}%
\makeatletter
\providecommand{\hrefCMSnoop }[0]{\@secondoftwo}%
\makeatother
\providecommand{\doi}{\texttt{doi:}\begingroup \urlstyle{tt}\Url}

\bibitem{PDG2024}
\hrefCMSnoop {}{{Particle Data Group}, S.~Navas { et~al.}, ``Review of particle
  physics'',} \textit{ Phys. Rev. D} \textbf{ 110} (2024) 030001,
  \href{http://dx.doi.org/10.1103/PhysRevD.110.030001}{\doi{10.1103/PhysRevD.110.030001}}.

\bibitem{HFLAV2024}
\hrefCMSnoop {}{{HFLAV} Collaboration, ``Averages of $b$-hadron, $c$-hadron,
  and $\tau$-lepton properties as of 2023'',} 2024.
  \href{http://www.arXiv.org/abs/2411.18639}{\texttt{arXiv:2411.18639}}.

\bibitem{Burdman:2001tf}
\hrefCMSnoop {}{G.~Burdman, E.~Golowich, J.~Hewett, and S.~Pakvasa, ``{Rare
  charm decays in the standard model and beyond}'',} \textit{ Phys. Rev. D}
  \textbf{ 66} (2002) 014009,
  \href{http://dx.doi.org/10.1103/PhysRevD.66.014009}{\doi{10.1103/PhysRevD.66.014009}},
  \href{http://www.arXiv.org/abs/hep-ph/0112235}{\texttt{arXiv:hep-ph/0112235}}.

\bibitem{PhysRevD.79.114030}
\hrefCMSnoop {}{E.~Golowich, J.~Hewett, S.~Pakvasa, and A.~A. Petrov,
  ``{Relating $\PDz$ - $\PADz$ mixing and $\PDz\to\ell^{+}\ell^{-}$ with new
  physics}'',} \textit{ Phys. Rev. D} \textbf{ 79} (2009) 114030,
  \href{http://dx.doi.org/10.1103/PhysRevD.79.114030}{\doi{10.1103/PhysRevD.79.114030}},
  \href{http://www.arXiv.org/abs/0903.2830}{\texttt{arXiv:0903.2830}}.

\bibitem{LQ3}
\hrefCMSnoop {}{S.~Fajfer and N.~Ko{\v{s}}nik, ``Leptoquarks in flavor changing
  neutral current charm decays'',} \textit{ Phys. Rev. D} \textbf{ 79} (2009)
  017502,
  \href{http://dx.doi.org/10.1103/PhysRevD.79.017502}{\doi{10.1103/PhysRevD.79.017502}},
  \href{http://www.arXiv.org/abs/0810.4858}{\texttt{arXiv:0810.4858}}.

\bibitem{LQ4}
I.~Dor{\v{s}}ner\hrefCMSnoop {}{ { et~al.}, ``{Physics of leptoquarks in
  precision experiments and at particle colliders}'',} \textit{ Phys. Rept.}
  \textbf{ 641} (2016) 1,
  \href{http://dx.doi.org/10.1016/j.physrep.2016.06.001}{\doi{10.1016/j.physrep.2016.06.001}},
  \href{http://www.arXiv.org/abs/1603.04993}{\texttt{arXiv:1603.04993}}.

\bibitem{eft1}
\hrefCMSnoop {}{S.~Fajfer and N.~Ko{\v{s}}nik, ``Prospects of discovering new
  physics in rare charm decays'',} \textit{ Eur. Phys. J. C} \textbf{ 75}
  (2015), no.~12, 567,
  \href{http://dx.doi.org/10.1140/epjc/s10052-015-3801-2}{\doi{10.1140/epjc/s10052-015-3801-2}},
  \href{http://www.arXiv.org/abs/1510.00965}{\texttt{arXiv:1510.00965}}.

\bibitem{eft2}
\hrefCMSnoop {}{A.~Bharucha, D.~Boito, and C.~M{\'e}aux, ``{Disentangling QCD
  and new physics in $D^+\to\pi^+\ell^+\ell^-$}'',} \textit{ JHEP} \textbf{ 04}
  (2021) 158,
  \href{http://dx.doi.org/10.1007/JHEP04(2021)158}{\doi{10.1007/JHEP04(2021)158}},
  \href{http://www.arXiv.org/abs/2011.12856}{\texttt{arXiv:2011.12856}}.

\bibitem{eft3}
\hrefCMSnoop {}{H.~Gisbert, G.~Hiller, and D.~Suelmann, ``{Effective field
  theory analysis of rare \abs{\Delta c} = \abs{\Delta u} = 1 charm decays}'',}
  \textit{ JHEP} \textbf{ 12} (2024) 102,
  \href{http://dx.doi.org/10.1007/JHEP12(2024)102}{\doi{10.1007/JHEP12(2024)102}},
  \href{http://www.arXiv.org/abs/2410.00115}{\texttt{arXiv:2410.00115}}.

\bibitem{LHCb:2022jaa}
\hrefCMSnoop {}{{LHCb Collaboration}, ``{Search for rare decays of \PDz mesons
  into two muons}'',} \textit{ Phys. Rev. Lett.} \textbf{ 131} (2023) 041804,
  \href{http://dx.doi.org/10.1103/PhysRevLett.131.041804}{\doi{10.1103/PhysRevLett.131.041804}},
  \href{http://www.arXiv.org/abs/2212.11203}{\texttt{arXiv:2212.11203}}.

\bibitem{parkingtrigger}
\hrefCMSnoop {}{{CMS Collaboration}, ``{Enriching the physics program of the
  CMS experiment via data scouting and data parking}'',} \textit{ Phys. Rep.}
  \textbf{ 1115} (2025) 678,
  \href{http://dx.doi.org/10.1016/j.physrep.2024.09.006}{\doi{10.1016/j.physrep.2024.09.006}},
  \href{http://www.arXiv.org/abs/2403.16134}{\texttt{arXiv:2403.16134}}.

\bibitem{hepdata}
\hrefCMSnoop {}{}{HEPD}ata record for this analysis, 2025.
\newblock
  \href{http://dx.doi.org/10.17182/hepdata.158634}{\doi{10.17182/hepdata.158634}}.

\bibitem{CMS:2008xjf}
\hrefCMSnoop {}{{CMS Collaboration}, ``The {CMS} experiment at the {CERN}
  {LHC}'',} \textit{ JINST} \textbf{ 3} (2008) S08004,
  \href{http://dx.doi.org/10.1088/1748-0221/3/08/S08004}{\doi{10.1088/1748-0221/3/08/S08004}}.

\bibitem{CMS:2023gfb}
\hrefCMSnoop {}{{CMS Collaboration}, ``Development of the {CMS} detector for
  the {CERN LHC Run 3}'',} \textit{ JINST} \textbf{ 19} (2024) P05064,
  \href{http://dx.doi.org/10.1088/1748-0221/19/05/P05064}{\doi{10.1088/1748-0221/19/05/P05064}}.

\bibitem{CMS:2020cmk}
\hrefCMSnoop {}{{CMS Collaboration}, ``Performance of the {CMS} {Level-1}
  trigger in proton-proton collisions at {$\sqrt{s} = 13\TeV$}'',} \textit{
  JINST} \textbf{ 15} (2020) P10017,
  \href{http://dx.doi.org/10.1088/1748-0221/15/10/P10017}{\doi{10.1088/1748-0221/15/10/P10017}},
  \href{http://www.arXiv.org/abs/2006.10165}{\texttt{arXiv:2006.10165}}.

\bibitem{CMS:2016ngn}
\hrefCMSnoop {}{{CMS Collaboration}, ``{The {CMS} trigger system}'',} \textit{
  JINST} \textbf{ 12} (2017) P01020,
  \href{http://dx.doi.org/10.1088/1748-0221/12/01/P01020}{\doi{10.1088/1748-0221/12/01/P01020}},
\href{http://www.arXiv.org/abs/1609.02366}{\texttt{arXiv:1609.02366}}.

\bibitem{CMS:2024aqx}
\hrefCMSnoop {}{{CMS Collaboration}, ``Performance of the {CMS} high-level
  trigger during {LHC} run 2'',} \textit{ JINST} \textbf{ 19} (2024) P11021,
  \href{http://dx.doi.org/10.1088/1748-0221/19/11/P11021}{\doi{10.1088/1748-0221/19/11/P11021}},
  \href{http://www.arXiv.org/abs/2410.17038}{\texttt{arXiv:2410.17038}}.

\bibitem{CMS:2020uim}
\hrefCMSnoop {}{{CMS Collaboration}, ``Electron and photon reconstruction and
  identification with the {CMS} experiment at the {CERN} {LHC}'',} \textit{
  JINST} \textbf{ 16} (2021) P05014,
  \href{http://dx.doi.org/10.1088/1748-0221/16/05/P05014}{\doi{10.1088/1748-0221/16/05/P05014}},
  \href{http://www.arXiv.org/abs/2012.06888}{\texttt{arXiv:2012.06888}}.

\bibitem{CMS:2018rym}
\hrefCMSnoop {}{{CMS Collaboration}, ``{Performance of the CMS muon detector
  and muon reconstruction with proton-proton collisions at $\sqrt{s}=$ 13
  TeV}'',} \textit{ JINST} \textbf{ 13} (2018) P06015,
  \href{http://dx.doi.org/10.1088/1748-0221/13/06/P06015}{\doi{10.1088/1748-0221/13/06/P06015}},
  \href{http://www.arXiv.org/abs/1804.04528}{\texttt{arXiv:1804.04528}}.

\bibitem{CMS:2014pgm}
\hrefCMSnoop {}{{CMS Collaboration}, ``Description and performance of track and
  primary-vertex reconstruction with the {CMS} tracker'',} \textit{ JINST}
  \textbf{ 9} (2014) P10009,
  \href{http://dx.doi.org/10.1088/1748-0221/9/10/P10009}{\doi{10.1088/1748-0221/9/10/P10009}},
  \href{http://www.arXiv.org/abs/1405.6569}{\texttt{arXiv:1405.6569}}.

\bibitem{CMS:2017yfk}
\hrefCMSnoop {}{{CMS Collaboration}, ``Particle-flow reconstruction and global
  event description with the {CMS} detector'',} \textit{ JINST} \textbf{ 12}
  (2017) P10003,
  \href{http://dx.doi.org/10.1088/1748-0221/12/10/P10003}{\doi{10.1088/1748-0221/12/10/P10003}},
\href{http://www.arXiv.org/abs/1706.04965}{\texttt{arXiv:1706.04965}}.

\bibitem{Sjostrand:2014zea}
T.~Sj{\"o}strand\hrefCMSnoop {}{ { et~al.}, ``{An introduction to {PYTHIA}
  8.2}'',} \textit{ Comput. Phys. Commun.} \textbf{ 191} (2015) 159,
  \href{http://dx.doi.org/10.1016/j.cpc.2015.01.024}{\doi{10.1016/j.cpc.2015.01.024}},
\href{http://www.arXiv.org/abs/1410.3012}{\texttt{arXiv:1410.3012}}.

\bibitem{CMS:2019csb}
\hrefCMSnoop {}{{CMS Collaboration}, ``Extraction and validation of a new set
  of {CMS} {PYTHIA8} tunes from underlying-event measurements'',} \textit{ Eur.
  Phys. J. C} \textbf{ 80} (2020) 4,
  \href{http://dx.doi.org/10.1140/epjc/s10052-019-7499-4}{\doi{10.1140/epjc/s10052-019-7499-4}},
  \href{http://www.arXiv.org/abs/1903.12179}{\texttt{arXiv:1903.12179}}.

\bibitem{Agostinelli:2002hh}
\hrefCMSnoop {}{{GEANT4} Collaboration, ``{\GEANTfour}---a simulation
  toolkit'',} \textit{ Nucl. Instrum. Meth. A} \textbf{ 506} (2003) 250,
\href{http://dx.doi.org/10.1016/S0168-9002(03)01368-8}{\doi{10.1016/S0168-9002(03)01368-8}}.

\bibitem{Lange:2001uf}
\hrefCMSnoop {}{D.~J. Lange, ``The {EvtGen} particle decay simulation
  package'',} \textit{ Nucl. Instrum. Meth. A} \textbf{ 462} (2001) 152,
\href{http://dx.doi.org/10.1016/S0168-9002(01)00089-4}{\doi{10.1016/S0168-9002(01)00089-4}}.

\bibitem{Davidson:2010ew}
\hrefCMSnoop {}{N.~Davidson, T.~Przedzinski, and Z.~Was, ``{PHOTOS interface in
  C++: technical and physics documentation}'',} \textit{ Comput. Phys. Commun.}
  \textbf{ 199} (2016) 86,
  \href{http://dx.doi.org/10.1016/j.cpc.2015.09.013}{\doi{10.1016/j.cpc.2015.09.013}},
\href{http://www.arXiv.org/abs/1011.0937}{\texttt{arXiv:1011.0937}}.

\bibitem{Prokofiev:2005zz}
\hrefCMSnoop {}{K.~Prokofiev and T.~Speer, ``{A kinematic and a decay chain
  reconstruction library}'',} in \textit{ {\it Proceedings of Computing in High
  Energy and Nuclear Physics 2004\/}}, p.~411.
\newblock 2005.
\newblock
  \href{http://dx.doi.org/10.5170/CERN-2005-002.411}{\doi{10.5170/CERN-2005-002.411}}.

\bibitem{Chen:2016:XST:2939672.2939785}
\hrefCMSnoop {}{T.~Chen and C.~Guestrin, ``{XGBoost}: A scalable tree boosting
  system'',} in \textit{ \it Proc. 22nd ACM SIGKDD Int. Conf. on Knowledge
  Discovery and Data Mining\/}, KDD '16, p.~785.
\newblock ACM, New York, NY, USA, 2016.
\newblock
  \href{http://dx.doi.org/10.1145/2939672.2939785}{\doi{10.1145/2939672.2939785}}.

\bibitem{kfold}
\hrefCMSnoop {}{A.~Blum, A.~Kalai, and J.~Langford, ``Beating the hold-out:
  Bounds for k-fold and progressive cross-validation'',} in \textit{ Proc. 12th
  Ann. Conf. on Computational Learning Theory}, COLT '99, p.~203.
\newblock 1999.
\newblock
  \href{http://dx.doi.org/10.1145/307400.307439}{\doi{10.1145/307400.307439}}.

\bibitem{ref:bernstein}
\hrefCMSnoop {}{S.~Bernstein, ``D\'{e}monstration du th\'{e}or\`{e}me de
  {W}eierstrass fond\'{e}e sur le calcul des probabiliti\'{e}s'',} \textit{
  Comm. Soc. Math. Kharkov} \textbf{ 13} (1912) 1.

\bibitem{Dauncey:2014xga}
\hrefCMSnoop {}{P.~D. Dauncey, M.~Kenzie, N.~Wardle, and G.~J. Davies,
  ``{Handling uncertainties in background shapes}: {the discrete profiling
  method}'',} \textit{ JINST} \textbf{ 10} (2015) P04015,
  \href{http://dx.doi.org/10.1088/1748-0221/10/04/P04015}{\doi{10.1088/1748-0221/10/04/P04015}},
  \href{http://www.arXiv.org/abs/1408.6865}{\texttt{arXiv:1408.6865}}.

\bibitem{ref:TnP}
\hrefCMSnoop {}{{CMS Collaboration}, ``{Performance of CMS muon reconstruction
  in pp collision events at $\sqrt{s} = 7$ TeV}'',} \textit{ JINST} \textbf{ 7}
  (2012) P10002,
  \href{http://dx.doi.org/10.1088/1748-0221/7/10/P10002}{\doi{10.1088/1748-0221/7/10/P10002}},
  \href{http://www.arXiv.org/abs/1206.4071}{\texttt{arXiv:1206.4071}}.

\bibitem{DP-2022-012}
\href {https://cds.cern.ch/record/2810814}{{CMS Collaboration}, ``{Tracking
  performances for charged pions with Run 2 legacy data}'',} CMS Detector
  Performance Note CMS-DP-2022-012, 2022.

\bibitem{combinetool}
\hrefCMSnoop {}{{CMS Collaboration}, ``{The CMS statistical analysis and
  combination tool: COMBINE}'',} \textit{ Comput. Softw. Big Sci.} \textbf{ 8}
  (2024) 19,
  \href{http://dx.doi.org/10.1007/s41781-024-00121-4}{\doi{10.1007/s41781-024-00121-4}},
  \href{http://www.arXiv.org/abs/2404.06614}{\texttt{arXiv:2404.06614}}.

\bibitem{Junk}
\hrefCMSnoop {}{T.~Junk, ``Confidence level computation for combining searches
  with small statistics'',} \textit{ Nucl. Instrum. Meth. A} \textbf{ 434}
  (1999) 435,
  \href{http://dx.doi.org/10.1016/S0168-9002(99)00498-2}{\doi{10.1016/S0168-9002(99)00498-2}},
  \href{http://www.arXiv.org/abs/hep-ex/9902006}{\texttt{arXiv:hep-ex/9902006}}.

\bibitem{Read}
\hrefCMSnoop {}{A.~L. Read, ``Presentation of search results: The {$\CLs$}
  technique'',} \textit{ J. Phys. G} \textbf{ 28} (2002) 2693,
  \href{http://dx.doi.org/10.1088/0954-3899/28/10/313}{\doi{10.1088/0954-3899/28/10/313}}.

\bibitem{CMS:softmuonid}
\hrefCMSnoop {}{{CMS Collaboration}, ``{Identification of low-momentum muons in
  the CMS detector using multivariate techniques in proton-proton collisions at
  $\sqrt{s}=$ 13.6 TeV}'',} \textit{ JINST} \textbf{ 20} (2025) P04021,
  \href{http://dx.doi.org/10.1088/1748-0221/20/04/P04021}{\doi{10.1088/1748-0221/20/04/P04021}},
  \href{http://www.arXiv.org/abs/2412.17590}{\texttt{arXiv:2412.17590}}.

\end{thebibliography}\endgroup

\appendix
\section{Details of \texorpdfstring{\PDz}{D0} normalization}\label{end_matter_a}
The normalizations of the signal, \dzmm, and backgrounds, \dzpipi and \dzpimunu, are derived with the yield of the normalization decay channel $\dzpipi$ measured from the zero bias data set: 
\ifthenelse{\boolean{cms@external}}
{ 
\begin{multline*}
  N_{\dstarmm} = N_{\dstarpipi} S_{\mathrm{ZB}}\frac{\varepsilon_{\dstarmm}}{\varepsilon_{\dstarpipi}} \frac{\mathcal{B}_{\dstarmm}}{\mathcal{B}_{\dstarpipi}} \\
  \times \mvadcorrection T_{\text{cor}},
\end{multline*}
\begin{multline*}
  N_{\dstarpipimm} = N_{\dstarpipi} S_{\mathrm{ZB}} \frac{\varepsilon_{\dstarpipimm}}{\varepsilon_{\dstarpipi}} \\
  \times \left(f_{\pi\to\mu}\right)^2 \mvadcorrection T_{\text{cor}},
\end{multline*}
\begin{multline*}
  N_{\dstarpimunu} = N_{\dstarpipi} S_{\mathrm{ZB}} \frac{\varepsilon_{\dstarpimunu}}{\varepsilon_{\dstarpipi}} \frac{\mathcal{B}_{\dstarpimunu}}{\mathcal{B}_{\dstarpipi}} \\
  \times \left(f_{\pi\to\mu}\right) T_{\text{cor}},
\end{multline*}
}
{
\begin{align*}
  N_{\dstarmm} &= N_{\dstarpipi} S_{\mathrm{ZB}}
  \, \frac{\varepsilon_{\dstarmm}}{\varepsilon_{\dstarpipi}}
  \, \frac{\mathcal{B}_{\dstarmm}}{\mathcal{B}_{\dstarpipi}}
  \, \mvadcorrection
  \, T_{\text{cor}},\\
  N_{\dstarpipimm} &= N_{\dstarpipi}\, S_{\mathrm{ZB}}
  \, \frac{\varepsilon_{\dstarpipimm}}{\varepsilon_{\dstarpipi}}
  \, \left(f_{\pi\to\mu}\right)^2
  \, \mvadcorrection
  \, T_{\text{cor}},\\
  N_{\dstarpimunu} &= N_{\dstarpipi}\, S_{\mathrm{ZB}}
  \, \frac{\varepsilon_{\dstarpimunu}}{\varepsilon_{\dstarpipi}}
  \, \frac{\mathcal{B}_{\dstarpimunu}}{\mathcal{B}_{\dstarpipi}}
  \, \left(f_{\pi\to\mu}\right)
  \, T_{\text{cor}},
\end{align*}
}
where $N_X$, $\varepsilon_X$, and $\mathcal{B}_X$ are the yield, efficiency extracted from simulation, and decay branching fraction for process $X$, respectively, $S_{\mathrm{ZB}}$ is the effective prescale factor for the zero bias data set, $f_{\pi\to\mu}$ is the pion to muon misidentification ratio between data and simulation, \mvadcorrection is the correction used to account for the difference in \mvad efficiency when applied to different channels, and $T_{\text{cor}}$ is the efficiency correction for the dimuon trigger. The background from \dzpipi decays is denoted as \dstarpipimm to distinguish it from the normalization channel. 

The efficiencies are obtained from simulation for the selected regions of \mmm and \dm. The misidentification of a pion as a muon can be separated into two components: decay-in-flight (DIF) in which the pion decays to a muon within the tracker and non-DIF. In the selected \mmm region, the relative contributions of two DIF muons, one of each, and two non-DIF muons for the \dzpipi process are 81\%, 18\%, 1\%, respectively, and for \dzpimunu the relative amounts are 90\% DIF and 10\% non-DIF, as determined from simulation. The simulated samples used to determine the shapes of these two processes were generated with the charged pion lifetime shortened by a factor of 50 to reduce computing requirements. The generated events are then reweighted using the lifetime of each pion in the \PDz decay to match the true pion lifetime distribution. This sample is intended to determine the DIF contribution. To assess the non-DIF contribution, simulated events with no lifetime modification are used, with the DIF contributions removed. The two shapes are combined based on the proportions found (without including the 1\% from non-DIF pairs) and the efficiency is evaluated. A systematic uncertainty is also assessed for the \dzpimunu process to account for the observation of a correlation between the \mmm value and the \dm width. As the \mmm approaches the true \PDz mass, the \dm width decreases since the momentum carried by the neutrino is reduced. A separate fit that includes a model of this correlation is used to determine the systematic uncertainty.

The muon misidentification rate is measured in data with \kspipi decays~\cite{CMS:softmuonid} and is found to be $1.3\times 10^{-3}$ per pion on average. This rate is also measured in simulation and compared to the data as a function of muon \pt and \PKzS decay location. No significant dependence on these variables is found and the measured scale factor is $f_{\pi\to\mu}=1.06$. The systematic uncertainty is evaluated in several ways. A 7\% uncertainty is assigned to the scale factor, accounting for statistical uncertainty, differences across data-taking periods, and variations among simulation samples.  Differences in meson decay location and muon \pt distributions between the \kspipi and \dzpipi processes contribute an additional 4.5\% uncertainty. The relative amount of DIF and non-DIF contributions is also varied, leading to a conservative 11\% uncertainty. The total systematic uncertainty is 14\%. 

The shapes of the background components with misidentified muons are validated using \kspipi decays, where both pions are misidentified as muons, and good agreement in the mass distribution is observed between data and simulation. The impact of possible mismodeling is found to be negligible.

The dimuon trigger efficiency is estimated from simulated events and is corrected in
data using high-efficiency reference triggers. The L1 trigger efficiency correction
is derived by measuring the efficiency of the dimuon L1 trigger relative to the single-muon triggers, which have efficiencies close to 100\% in events containing two muons. The ratio of data-to-simulation efficiencies is $(99.0\pm0.5)\%$, with the uncertainty including a $0.3\%$ statistical uncertainty and a $0.4\%$ variations in the ratio across different data-taking periods and single muon triggers.
The HLT efficiency correction is derived by
measuring the efficiency of the dimuon HLT with respect to
the L1 dimuon trigger. The ratio of data-to-simulation efficiencies is $(95.3\pm0.5)\%$, 
with the uncertainty including a $0.2\%$ statistical uncertainty and a $0.4\%$ variation across different data-taking periods. 
The total correction is estimated as the product of the L1 and HLT corrections and is $(94.3\pm0.7)$\%.  

The \mvad correction is determined from simulation, where the difference in efficiency between \dzmm, \dzpipi, \dzpipimm, and \dzpimunu components is within 2\%.  The uncertainty is taken as the sum in quadrature of the statistical uncertainty and the size of the correction.
A discrepancy in the \mvad distribution between data and simulation is observed in all of the \PDz decays. A correction for this difference was made by reweighting the \mvad distribution of the simulated events to match the data, using \dzkpi decays. The effect on the relevant efficiency ratios was negligible ($0.3\%$) and therefore this correction was not applied, and no systematic uncertainty was assessed.

As the \mmm and \dm shapes for the \dzpipimm are similar to the signal, it is important to check the yield determination. This is done by utilizing \dzkpi decays, which have a branching fraction 27 times larger than \dzpipi. For this process, the kaon-to-muon misidentification rate is needed. It is measured in data from \phikk decays to be $2.7 \times 10^{-3}$ per kaon. The observed yield of the \dzkpi process in which both charged hadrons are misidentified as muons is found to be consistent with the expected yield after scaling by the corresponding efficiencies and branching fraction.

The event yields for each component of the fit are summarized in Table~\ref{tab:unblind_yields_table}. 

\begin{table*}[hptb!]
  \centering
  \topcaption{The post-fit event yields for the signal, the combinatorial background, the \dzpipi background, and the \dzpimunu background. The observed numbers of events are given in the Data column. The subrange is in one dimension with a full range in the other dimension.}
    \cmsTable{
    \begin{scotch}{lcccccc}
       Range & Signal  & Combinatorial background & $\dzpipi$ & $\dzpimunu$  & Data\\
      \hline
 Full range & $100\pm 120$ & $126\,140\pm 380$ & $278\pm 51$ & $231\pm 40$ & 126\,752 \\
 $0.145<\dm<0.146\GeV$ & $67\pm 81$ & $14\,037\pm 42 $ & $179\pm 33$ & $94\pm 16$ & 14\,412 \\
 $1.84<\mmm<1.89\GeV$ & $90\pm 110$ & $48\,530\pm 150$ & $162\pm 30$ & $62\pm 11$ & 48\,798 \\
    \end{scotch}
    }
  \label{tab:unblind_yields_table}
\end{table*}
\cleardoublepage \section{The CMS Collaboration \label{app:collab}}\begin{sloppypar}\hyphenpenalty=5000\widowpenalty=500\clubpenalty=5000
\cmsinstitute{Yerevan Physics Institute, Yerevan, Armenia}
{\tolerance=6000
V.~Chekhovsky, A.~Hayrapetyan, V.~Makarenko\cmsorcid{0000-0002-8406-8605}, A.~Tumasyan\cmsAuthorMark{1}\cmsorcid{0009-0000-0684-6742}
\par}
\cmsinstitute{Institut f\"{u}r Hochenergiephysik, Vienna, Austria}
{\tolerance=6000
W.~Adam\cmsorcid{0000-0001-9099-4341}, J.W.~Andrejkovic, L.~Benato\cmsorcid{0000-0001-5135-7489}, T.~Bergauer\cmsorcid{0000-0002-5786-0293}, S.~Chatterjee\cmsorcid{0000-0003-2660-0349}, K.~Damanakis\cmsorcid{0000-0001-5389-2872}, M.~Dragicevic\cmsorcid{0000-0003-1967-6783}, P.S.~Hussain\cmsorcid{0000-0002-4825-5278}, M.~Jeitler\cmsAuthorMark{2}\cmsorcid{0000-0002-5141-9560}, N.~Krammer\cmsorcid{0000-0002-0548-0985}, A.~Li\cmsorcid{0000-0002-4547-116X}, D.~Liko\cmsorcid{0000-0002-3380-473X}, I.~Mikulec\cmsorcid{0000-0003-0385-2746}, J.~Schieck\cmsAuthorMark{2}\cmsorcid{0000-0002-1058-8093}, R.~Sch\"{o}fbeck\cmsAuthorMark{2}\cmsorcid{0000-0002-2332-8784}, D.~Schwarz\cmsorcid{0000-0002-3821-7331}, M.~Sonawane\cmsorcid{0000-0003-0510-7010}, W.~Waltenberger\cmsorcid{0000-0002-6215-7228}, C.-E.~Wulz\cmsAuthorMark{2}\cmsorcid{0000-0001-9226-5812}
% ...
\par}
\vskip\cmsinstskip
\dag:~Deceased\\
$^{1}$Also at Yerevan State University, Yerevan, Armenia\\
$^{2}$Also at TU Wien, Vienna, Austria\\
% ...
\end{sloppypar}
\end{document}